\documentclass[11pt,twoside]{article}

\usepackage{asp2006}
\usepackage{epsf}
\usepackage{lscape}
\usepackage{graphicx}
\usepackage{natbib}

\begin{document}
\title{Numerical Models of Sgr A*}   
\author{Mo{\'s}cibrodzka M.\altaffilmark{1}, Gammie C.F.\altaffilmark{1,2},Dolence
J.\altaffilmark{2}, Shiokawa H.\altaffilmark{2}, Leung P.K.\altaffilmark{2}}   
\affil{$^1$Department of Physics, University of Illinois, 1110 West Green Street, Urbana, IL 61801\\
$^2$Astronomy Department, University of Illinois, 1002 West Green Street,
Urbana, IL 61801 }

\begin{abstract} 

We review results from general relativistic axisymmetric
magnetohydrodynamic simulations of accretion in Sgr A*.  We use
general relativistic radiative transfer methods and to produce a broad band
(from millimeter to gamma-rays) spectrum. Using a ray tracing scheme we also
model images of Sgr A* and compare the size of image to the VLBI
observations at 230 GHz. We perform a parameter survey and study
radiative properties of the flow models for various black hole spins,
ion to electron temperature ratios, and inclinations.  We scale our
models to reconstruct the flux and the spectral slope around 230 GHz.
The combination of Monte Carlo spectral energy distribution calculations
and 230 GHz image modeling constrains the parameter space of the
numerical models. Our models suggest rather high black hole spin
($a_*\approx 0.9$), electron temperatures close to the ion temperature
($T_i/T_e \sim 3$) and high inclination angles ($i \approx 90 \deg$).

\end{abstract}


\section{Introduction}   

Observations of the Galactic Center provide strong evidence for the
existence of a supermassive black hole in Sgr A* (which hereafter refers
to the black hole, the accretion flow, and the radio source).  Sgr A*'s
proximity allows us to perform observations with higher angular
resolution than other galactic nuclei.  Estimates of Sgr A*'s mass
$M=4.5 \pm 0.4 \times 10^6 M_{\odot}$ and distance $D = 8.4 \pm 0.4$ kpc
(\citealt{ghez:2008}, \citealt{gillessen:2009}) indicate that it has the
largest angular size of any known black hole ($G M/(c^2 D) \simeq 5.3
{\rm \mu as}$). 

Recent 230 GHz VLBI constrains the structure of Sgr A* on angular scales
comparable to the size of the black hole horizon
(\citealt{doeleman:2008}, see also Doeleman contribution to this
conference proceeding). Using a two-parameter, symmetric Gaussian
brightness distribution model VLBI infers a full width at half maximum
FWHM $= 37^{+16}_{-10}$ ${\rm \mu as}$.  This is smaller than the
apparent diameter of the black hole: $\approx 2 \sqrt{27} G M /(c^2 D)
\simeq 55 {\rm \mu as}$ (this depends only weakly on black hole spin).
Sgr~A* radio - submm emission is usually modeled as synchrotron
emission, with the turnover at $\sim 230$ GHz indicating transition from
optically thick to optically thin emission.  The model of accretion and
its geometry is still under debate because Sgr A* is dim at shorter
wavelengths (in NIR and X-ray band) or completely obscured (UV and
optical light). Moreover in the NIR and X-ray the source is not resolved
and we only have upper limits for its quiescent luminosity.

Sgr~A* is a strongly sub-Eddington source ($L_{Bol} \approx 10^{-9}
L_{Edd}$).  Models suggest that the source is accreting inefficiently
(in the sense that $L_{bol}/(\dot{M} c^2) \ll 1$), which justifies
treating the dynamics of the flow and its radiative properties
independently. Therefore most of Sgr~A* models published to this day
consists of two separate parts: plasma dynamics model and/or radiative
transfer model. We can further categorize the dynamical models into:
accretion flow vs. outflow models, stationary vs. time-dependent,
Newtonian/post-Newtonian vs. General Relativistic, models covering small
region (a couple of gravitational radii) vs.  large region (thousands of
gravitational radii). The radiative transfer modeling usually uses ray
tracing (always relativistic) or Monte Carlo methods (nonrelativistic as
well as fully relativistic).  Ray tracing allows one to model source
images at sub-mm frequencies and the SED of direct synchrotron emission
(radio,sub-mm).  Monte Carlo allows one to model the Compton scattering
and multiwavelength (from radio to gamma-rays) SED of Sgr~A*.  In
Table~\ref{table_models} we summarize the recent progress in models.

\begin{table}[!ht]
\smallskip
\begin{center}
{\small
\begin{tabular}{lcccc}
\tableline
Reference & dynamical & radiative & plasma& range  \\
          & model & model & & of model  \\
\tableline
\citet{narayan:1998} & stat. rel. ADAF & non-rel. MC & th & $10^5 R_g$\\
\citet{markoff:2001}& Jet& scaling & non-th & --\\
\citet{yuan:2003}& stat non-rel. RIAF & non-rel rays  & th+non-th& $2 \times 10^5 R_g$\\
\citet{ohsuga:2005}& MHD-time dep. & non-rel. MC & th & $60 R_g$\\
\citet{goldston:2005}& MHD-time dep. & polarized non-rel. rays & th+non-th & $512 R_g$\\
\citet{broderick:2006b}& stat. non-rel RIAF& polarized RT & non-th & $2 \times 10^5 R_g$\\
\citet{moscibrodzka:2007}& MHD-time dep. & non-rel. MC & th+non-th   & $2.4 \times 10^3 R_g$\\
\citet{loeb:2007}& Jet & scaling & th+non-th & --\\
\citet{huang:2007}& stat. RIAF& RT & th & $2 \times 10^5 R_g$ \\
\citet{markoff:2007}& Jet& non-rel rays /w corr & non-th & --\\
\citet{huang:2009}& stat. rel. RIAF & RT & th & $10^4 R_g$\\
\citet{broderick:2009}& stat. rel. RIAF & RT & th+non-th & $2 \times 10^5 R_g$ \\
\citet{chan:2009}& MHD-time dep.& non-rel rays /w corr. & th+non-th& $43 R_g$\\
\citet{yuan:2009}& stat. rel. RIAF& RT& th & $100 R_g$\\
\citet{hilburn:2009}& GRMHD-time dep.   & non-rel MC & th  & $40 R_g$\\
\citet{dexter:2009}& GRMHD-time dep.    & RT     & th   & $40 R_g$\\
\citet{moscibrodzka:2009}& GRMHD-time dep. & RT + rel. MC & th   & $40R_g$\\
\tableline
\end{tabular}
}
\caption{Summary of selected models of Sgr~A*. Abbreviations: RT-ray tracing,
MC-Monte Carlo, GR-general relativistic, RIAF-radiatively inefficient
accretion flow, ADAF-advection dominate accretion flow, plasma-particles
distribution, 
th-thermal, non-th-non-thermal, range-model radial range.
}\label{table_models}
\end{center}
\end{table}

\section{Methodology}

Our numerical model of accretion onto Sgr A* consist of: a physical
model of the accretion flow dynamics with its numerical realization, and
a radiative transfer model.  We assume that the accreting plasma is
geometrically thick, optically thin, turbulent and it accretes onto a
spinning black hole. The spin angular momentum $J$ of the black hole,
whose magnitude is parameterized by $a_* = J c/(G M^2)$, is assumed to
be aligned with the angular momentum of the accretion flow.

The accretion flow in Sgr~A* is collisionless.  We assume that ions and
electrons have thermal distributions, possibly with different
temperatures.  In our model we allow the electron temperature $T_e$ to
differ from the ion temperature $T_i$, but we fix the ratio
($T_i/T_e=const$).  The plasma equation of state is described by the
$P=e(\gamma-1)$ relation with $\gamma=13/9$ (non-relativistic ions and
relativistic electrons).  A more physical model would evolve $T_e$ and
$T_i$ independently with a model for dissipation and energy exchange
between the electrons and ions, but this would complicate the model and
introduce a host of new parameters.

We realize the physical model using an axisymmetric version of the GRMHD
code {\tt harm} (\citealt{gammie:2003}). As initial conditions we adopt
an analytical model of a thick disk (a torus) in hydrostatic equilibrium
\citep{fishbone:1976}.  Since the code is not designed to evolve MHD
equations in vacuum, we surround the equilibrium torus with a hot, low
density plasma which does not influence the torus evolution.  We seed
the torus with a poloidal, concentric loops of weak magnetic field.
Small perturbations are added to the internal energy which allows to for
magnetorotational instability and turbulence development. We solve the
GRMHD evolution equations of evolution until a quasi-equilibrium
accretion flow is established (meaning that the flow is not evolving on
the dynamical timescale). The details of torus initial setup and
discussion of the flow evolution is presented in \citet{mckinney:2004}.

Our numerical domain extends from the black hole event horizon to $40
GM/c^2 = 1.8$ AU or $210 \mu as$, but is only in equilibrium to $\sim 15
GM/c^2 = 0.7$AU or $80 \mu as$.  Since low frequency emission from Sgr
A* is believed to arise at larger radius, we are unable to model the low
frequency (radio, mm) portion of spectral energy distribution (SED).
Our (untested) hypothesis is that at $r < 15 GM/c^2$ the model
accurately represent the inner portions of a relaxed accretion flow
extending over many decades in radius.

Our fluid dynamical model is scale free but the radiative transfer
calculation is not.  Therefore we need to specify the simulation length
unit ${\mathcal L}=GM/c^2$, time unit ${\mathcal T}=GM/c^3$, and mass
unit ${\mathcal M}$ (which scales the mass accretion rate).  ${\mathcal
L}$ and ${\mathcal T}$ are defined by the black hole mass adopted from
observations.  ${\mathcal M}$ is a free parameter set to reproduce the
submillimeter flux $F_{230GHz}=3.4 Jy$, \citep{marrone:2006b}.  To
``observe'' the numerical model we need to specify the inclination $i$
of the black hole spin to the line of sight.  The distance to the
observer is assumed to be $D=8.4 kpc$.

\begin{figure}[!ht]
\begin{center}
\includegraphics[scale = 0.5, angle = 0]{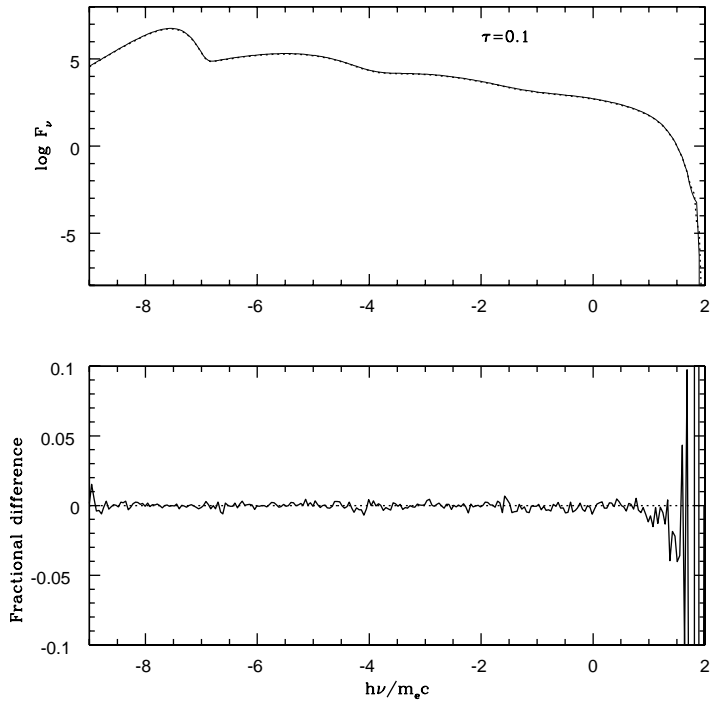}
\end{center}
\caption{Code testing: {\tt grmonty} (points) and {\tt Sphere} (independent
Monte Carlo code for radiative transfer, solid line) SED comparison.  In this
test problem a thermal point source emits photons from the center of the
spherical, hot plasma cloud. The cloud is characterized by two parameters:
optical thickness $\tau$ (here $\tau=0.1$) and electron temperature
$\Theta_e$ ($\Theta_e=kT_e/m_ec^2=4.$). Lower panel shows the difference
between two SEDs.
}\label{test_fig}
\end{figure}

The SED is calculated in a Monte Carlo fashion.  We use {\tt grmonty} - a
general relativistic radiative transfer scheme (\citealt{dolence:2009}) to
calculate the light propagation in the strong gravity taking into account
synchrotron emission and absorption and Compton scattering.  {\tt grmonty} has
been extensively tested. As one of our tests we compared the code performance
in a flat space with independent Monte Carlo scheme {\tt Sphere} (kindly
provided by its author, Shane Davies). In Fig.\ref{test_fig} we present a
standard test for Compton scattering in the spherical, homogenous, cloud of
hot plasma \citep{pss:1979}. 

In the present models we compute spectra through a time slice of a
simulation as if the light had infinite speed ('fast light'
approximation).  In reality the light crossing time is comparable to the
dynamical time which suggest radiative transfer through the changing
medium ($t_{dyn} \sim t_{cross} \sim 20$s in Sgr A*). Finally we use a
ray-tracing scheme, {\tt ibothros} (\citealt{noble:2007}), which
accounts only for synchrotron emission and absorption in order to
calculate the 230 GHz intensity maps of an accretion flow as seen by a
distant observer (again in the fast light approximation).  To estimate
the size of the emitting region we calculate the eigenvalues of the
matrix formed by taking the second angular moments of the image one the
sky (the lengths of the principal axes). The eigenvalues along the major
and minor axis are then compared to the FWHM VLBI constraint.

We accept or reject numerical models based on several observational
constraints: the flux at 230 GHz (we scale our models to reconstruct the
flux F=3.4 Jy at 230 GHz), submillimeter spectral slope $\alpha$ around
a turn over frequency (230-690 GHz, $\alpha$ changing from -0.46 to 0.08
\citealt{marrone:2006b}), upper limit for near-IR quiescent luminosity
(e.g. \citealt{genzel:2003}), an upper limit for X-ray luminosity $L_X <
2.4 \times 10^{33} [erg s^{-1}]$ ({\citealt{baganoff:2003}), size of the
image at 230 GHz ({\citealt{doeleman:2008}). We compare only
time-averaged SEDs and time-averaged images to the observations.

\section{Results}

\subsection*{Best-bet model}

We vary three (free) model parameters: spin of the black hole ($a_*$=0.5,
0.75, 0.84, 0.94, 0.96, 0.98), the ion to electron temperature ratio
($T_i/T_e=1, 3, 10$), and inclination of the observer with respect to
the spin axis ($i=5^{\deg}, 45^{\deg}$, and $85^{\deg}$). In
Table~\ref{tab1} we show a set of accretion flow models in which SEDs
are consistent with multiwavelength data as observed at at least one
inclinations.  We find the ``best-bet'' model, that best satisfies all
the observational constraints (including VLBI 230 GHz size constrains)
has spin $a_*=0.94$, $T_i/T_e=3$ and $i=85^{\deg}$.  All other models
are inconsistent with the observed SED or only marginally consistent
with the VLBI size measurement. 

\begin{table}[!ht]
\smallskip
\begin{center}
{\small
\begin{tabular}{cccccccc ccc}
\tableline
run & $a_*$ & $i$   &$ < \dot{M}\cdot 10^{-9}>$ & $\alpha$ &    $\log_{10} L_X$
&$\eta$&cons. \\ 
    &       & [deg] &$[M_{\odot}/yr^{-1}]$&& $[erg s^{-1}]$  &&w/
    obs.?
 
\\
\tableline
  &        &            5 & 1.9 & -1.68& 31.5 & $3.5 \times 10^{-2}$&  NO\\
D3 & 0.94   &            45& 1.7 & -1.27& 31.8 & $3.1 \times 10^{-2}$& NO\\
  &        &            85& 1.86 & -0.44& 32.9& $3.4 \times 10^{-2}$&  YES\\
\\
  &        &            5 & 90.8 & -1.37& 30.1 &$5.4 \times 10^{-4}$&    NO\\ 
A10 & 0.5    &            45& 117.1& -0.2 & 31.4 &$6.7 \times 10^{-4}$&YES\\ 
  &        &            85& 369.0& 1.38 & 33.5 &$1.6 \times 10^{-3}$&  NO\\ 
\\
  &        &            5 & 38.3 & -1.05& 30.0 &$1.6 \times 10^{-3}$&  NO\\
B10 & 0.75   &            45& 50.2 & 0.04 & 31.8 &$2.0 \times 10^{-3}$&YES\\
  &        &            85& 190.6& 1.49 & 34.6 &$6.9 \times 10^{-3}$&  NO\\
\\
  &        &            5 & 19.5 & -1.15& 31.4 &$5.1 \times 10^{-3}$&  NO\\
C10 & 0.875  &            45& 23.6 & -0.07& 32.0 &$6.2 \times 10^{-3}$&YES\\
  &        &            85& 41.4 & 1.19 & 34.2 &$1.7 \times 10^{-2}$&  NO\\
\\
  &        &            5 & 13.7 & -0.93 & 31.8 &$1.1 \times 10^{-2}$&  NO\\
D10 & 0.94   &            45& 15.2 & -0.05 & 32.3 &$1.1 \times 10^{-2}$&YES\\
  &        &            85& 31.2 & 1.17  & 34.4 &$2.5 \times 10^{-2}$&  NO\\
\\
  &        &            5 & 13.6 & -0.40 & 32.5 &$2.6 \times 10^{-2}$&  YES\\
E10 & 0.97   &            45& 14.3 & 0.2   & 33.1 &$2.8 \times 10^{-2}$&NO\\
  &        &            85& 26.5 & 1.19  & 35.2 &$5.1 \times 10^{-2}$&  NO\\
\\
  &        &            5 & 9.07 & -0.6  & 32.7 &$5.3 \times 10^{-2}$&  YES\\
F10 &0.98    &            45& 9.16 & -0.08 & 33.2 &$5.2 \times 10^{-2}$&NO\\
  &        &            85& 15.5 & 1.12  & 35.4 &$9.0 \times 10^{-2}$&  NO\\
\tableline
\end{tabular}
}
\caption{Chosen set of models with temperature ratio 3 and 10 which SEDs are
consistent with observations at least at one observation angle. The columns from left
to right are: run ID (the number is the run name indicates $T_i/T_e$),
dimensionless spin of the black hole, inclination angle of the observer with
respect to the black hole spin axis, averaged rest mass accretion rate,
$\alpha$ spectral slope between 230-690 GHz ($F\sim\nu^{\alpha}$), and
luminosity in the X-rays (at $\nu \sim 10^{18}$ Hz), the radiative efficiency
$\eta=L_{\rm BOL} / \dot{M} c^2$, and whether the model is consistent with the
data. We do not show models with $T_i/T_e=1$ here because they are all
inconsistent with the observations.}\label{tab1}
\end{center}
\end{table}

In Figure~\ref{fig1} we show the SED of our best-bet model. The SED
peaks around 690 GHz due to thermal synchrotron emission. Below 100 GHz
it fails to fit the data because that emission is produced outside the
simulation volume.  A second peak in the far UV is due to the first
Compton scattering order, at higher energies the photons are produced as
an effect of two or more scatterings. 

\begin{figure}[!ht]
\begin{center}
\includegraphics[clip,scale = 0.3, angle = 0]{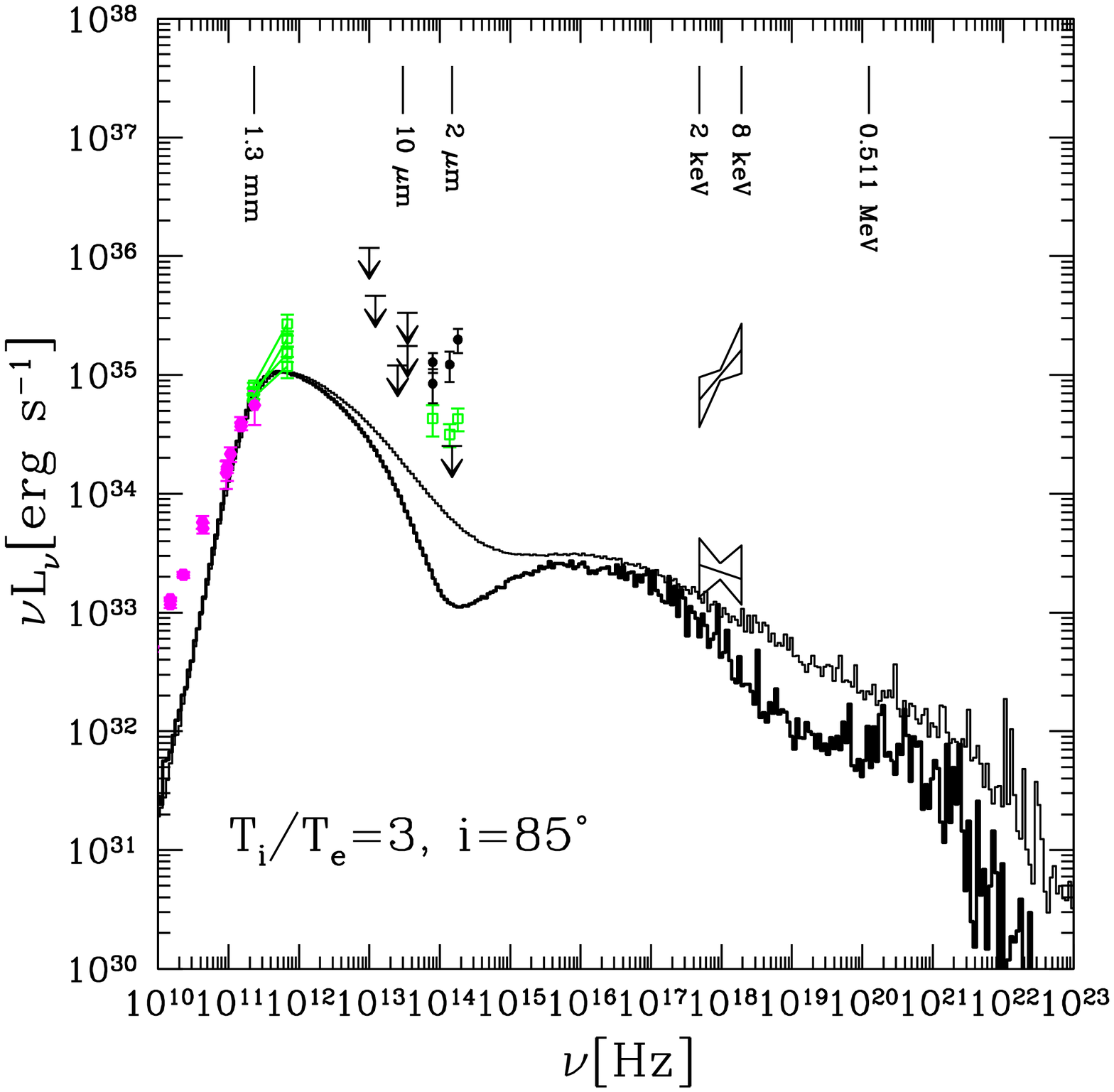} 
\includegraphics[clip,scale = 0.3, angle = -90, origin=c]{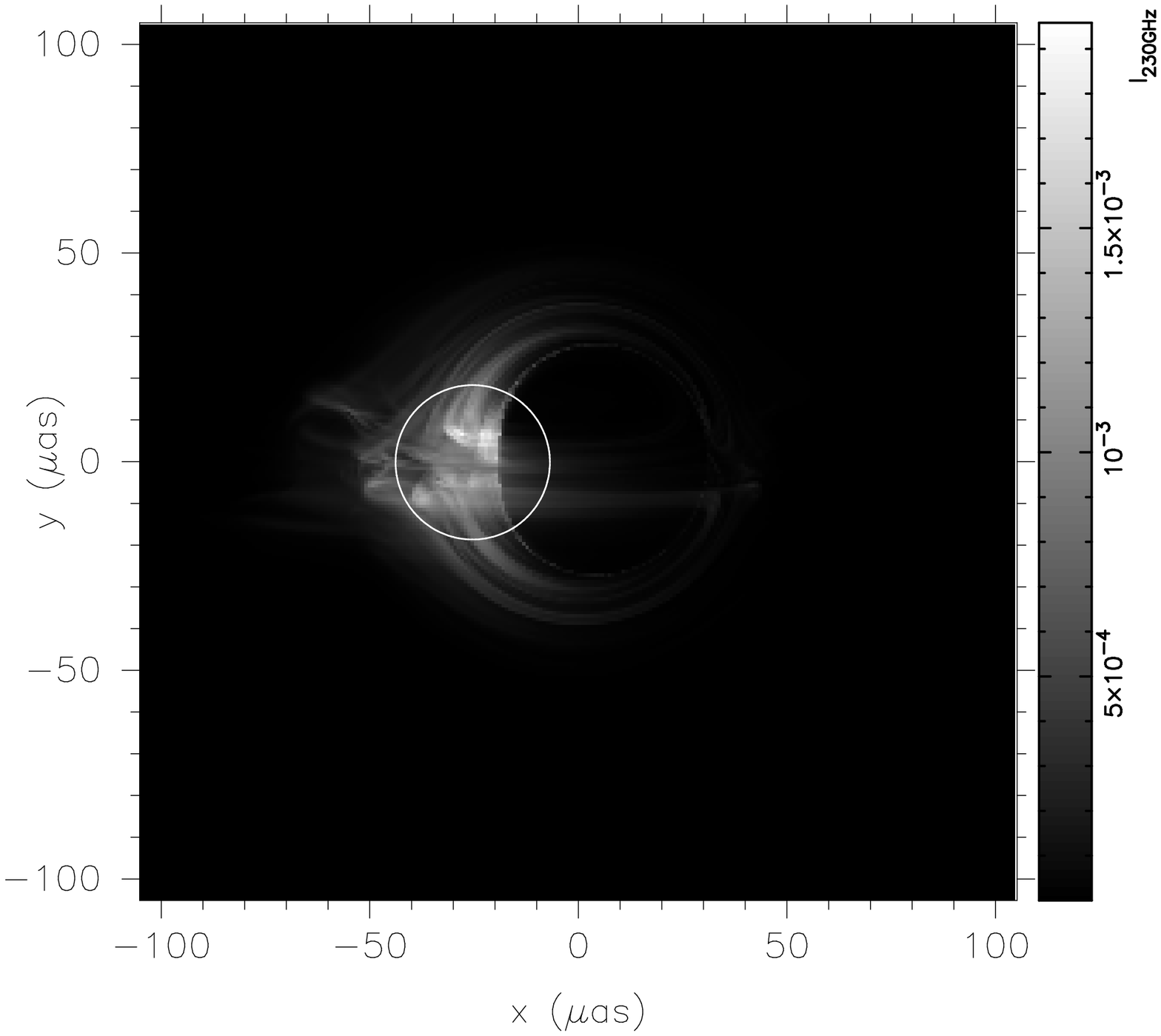}
\end{center}
\caption{
Left panel: SEDs computed based on a single time slice $t=1680 G M/c^3$ (thick
line) (see Figure~\ref{fig2} for the distributions of the physical variables
corresponding to the same time) along with the time averaged spectrum (thin
line) of our best-bet model.  Observational points are taken from:
\citealt{falcke:1998}, \citealt{an:2005}, \citealt{marrone:2006a} at radio,
\citealt{genzel:2003} at NIR (1.65, 2.16, and 3.76 ${\rm \mu m}$) and
\citealt{baganoff:2003} at X-rays (2-8 keV). The upper limits in the NIR band
are taken from \citealt{melia:2001} (30, 24.5 and 8.6 ${\rm \mu m}$),
\citealt{schoedel:2007} (8.6 $\mu m$) and \citealt{hornstein:2007} (2 ${\rm
\mu m}$). The points in the NIR at flaring state are from
\citealt{genzel:2003} (1.65, 2.16, and 3.76 ${\rm \mu m}$) , and
\citealt{dodds:2009} (3.8 ${\rm \mu m}$). An example of X-ray flare ($L_X = 1
\times 10^{35}$ $erg s^{-1}$) is taken from \citealt{baganoff:2001}. Right
panel: corresponding to SED in the left panel Image of the accretion flow at
$230 GHz$ correspindign to $t=1680 G M/c^3$.
Intensities are given in units of $erg s^{-1} {\rm
pixelsize^{-2} Hz^{-1} sr^{-1}}$, where the pixel size is $0.82 \mu as$.  The
image shows inner $40 GM/c^2$.  The white circle marks FWHM=37 $\mu as$ of a
symmetric Gaussian brightness profile centered at the image centroid.
}\label{fig1}
\end{figure}

In Figure~\ref{fig2}, we show the dynamical model corresponding to our
best-bet model.  We show maps of number density density, magnetic field
strength and electron temperature ($\theta_e=k_b T_e/m_ec^2$) on a
single time slice.
\begin{figure}[!ht]
\begin{center}
\includegraphics[scale = 0.6, angle = -90]{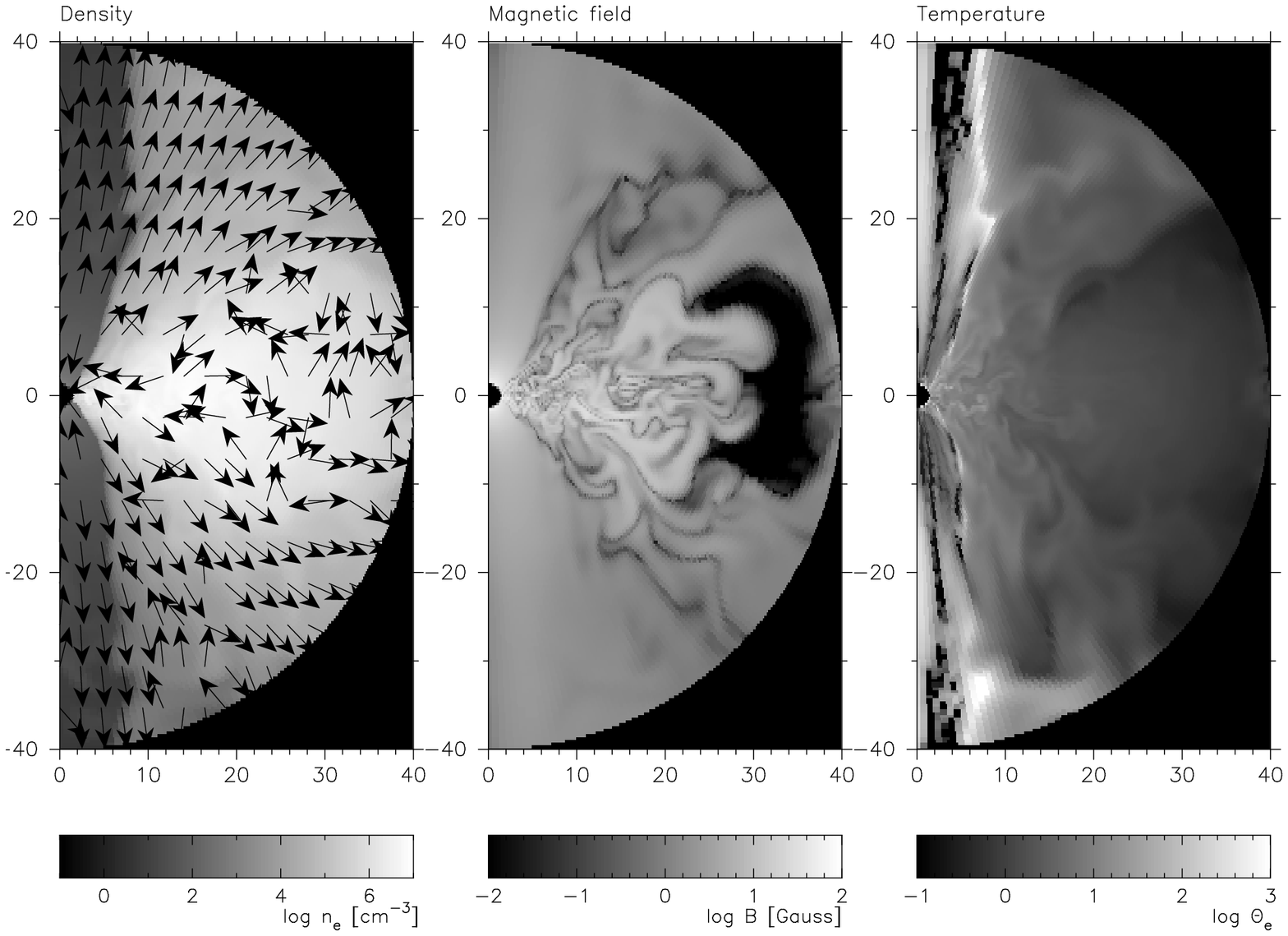}
\end{center}
\caption{
Accretion flow structure in our best-bet model with $a_* = 0.94$ and with
$T_i/T_e=3$ (model D3). The number density overplotted with the velocity
field, 
the magnetic field strength, and
the electron temperature are shown in the left, middle, and right panel
respectively. The axis units are $G M/c^2$.The figure shows a single 
time slice.}\label{fig2}
\end{figure}

Figure~\ref{fig3} maps the points of origin for photons below and in the
synchrotron peak (100-690 GHz), in the NIR ($10^{13}-10^{14} Hz$), and in the
X-ray (2-8 keV). The figure again corresponds to a single time slice
from our best-bet model presented in Figures~\ref{fig1},~\ref{fig2}.
Most of the submillimeter emission originates near the mid plane at $4 <
r c^2/(G M) < 6$. NIR photons are produced in the hot regions close to
the innermost circular orbit ($r_{ISCO}\approx 2.04 G M/c^2$). Photons
in the X-ray band are produces mainly by scatterings in the hottest
parts of the disk also close to the $r_{ISCO}$. A small fraction of
photons are emitted from the funnel wall at large radii ($15-40
GM/c^2$).  

\begin{figure}[!ht]
\begin{center}
\includegraphics[scale = 0.6, angle = -90]{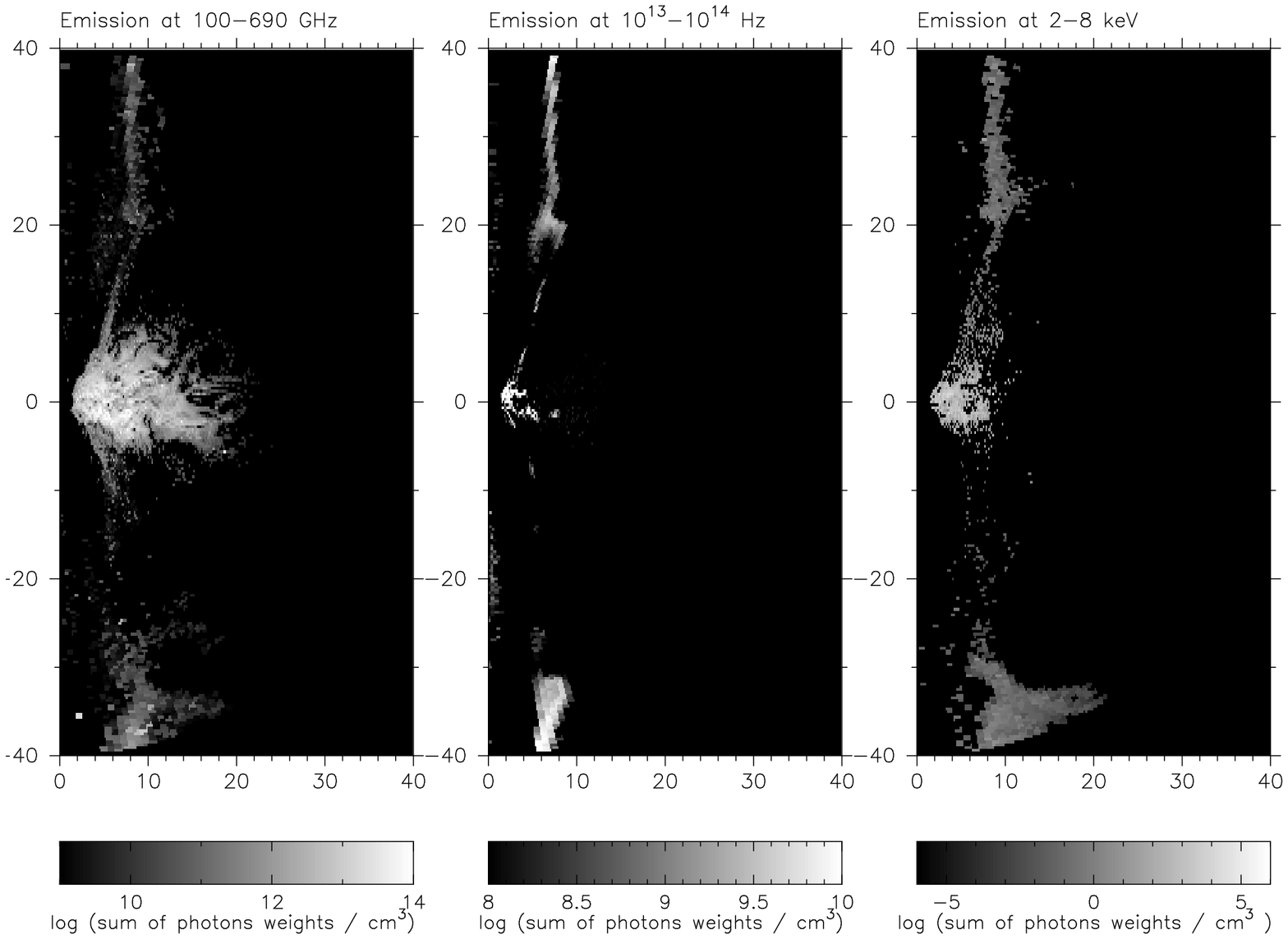}
\end{center}
\caption{
Maps showing the point of origin for photons in our best-bet model.  We show
the logarithm of the sum of the photon weights in each zone, which is
proportional to the number of photons seen at $100-690 GHz$ (left),
$10^{13}-10^{14} Hz$ (middle) and 2-8 keV (right) band. The axis scale units are
$GM/c^2$.
The figure presents a single time slice at $t=1680 G M/c^3$.  Note that gray
scale bands differ in scales.}\label{fig3}
\end{figure}

\subsection*{Summary of parameter survey}

We can draw the following general conclusions from our work: (1) Very
few of the time averaged SEDs based on a single-temperature
($T_i/T_e=1$) models produce the correct spectral slope between 230-690
GHz.  The exception is edge-on tori ($i=85^{\deg}$) around a fast
spinning black hole ($a_*=0.98,0.96$), but these models overproduce NIR
and X-ray flux. 

(2) For $T_i/T_e=3$ the only one model with $a_*=0.94$ at $i=85^{\deg}$
is consistent with all observational constraints. For $i=85\deg$, models
with spins below $a_* = 0.94$ are ruled out by the inconsistent spectral
slope, and models with higher spins ($a_*>0.94$), although consistent
with the observed $\alpha$, overproduce the quiescent NIR and X-ray
emission.  All models with $T_i/T_e=3$ observed at $i=5^{\deg}$ and
$45^{\deg}$ are ruled out by the inconsistent $\alpha$.

(3) For $T_i/T_e=10$, we find that all models with $i=85^{\deg}$ are
ruled out by both incorrect $\alpha$ and violation of NIR and X-ray
limits.  For lower inclination angles ($i=5^{\deg},45^{\deg}$) a few
models (E10 and F10 with $i = 5\deg$, and A10, B10, C10, and D10 at $i =
45\deg$) reproduce the observed $\alpha$. These models are consistent
with X-rays and NIR limitations.  Models E and F for $i=45\deg$ are
ruled out by NIR and X-ray limitations whereas models A10, B10, C10 and
D10 for $i=5\deg$ produce $\alpha$ which is too small.

(4) The dependence on $a_*$ arises largely because as $a_*$ increases
the inner edge of the disk --- the ISCO -- reaches deeper into the
gravitational potential of the black hole, where the temperature and
magnetic field strength are higher.  In the disk mid-plane, the
temperature is proportional to the virial temperature and scales with
radius $\Theta_e \propto 1/r$. $B \propto 1/r$, while the density $\sim
r$, below the pressure maximum (longer simulations in 3D show a more
relaxed, declining density profile).  Holding all else constant this
implies a higher peak frequency for synchrotron emission, a constant
Thomson depth (in our models the Thomson depth at the ISCO is roughly
constant, since the path length $1/\sim r_{ISCO}$ but the density $\sim
r_{ISCO}$), and a larger energy boost per scattering $A \approx 16
\Theta_e^2$, as can be seen in comparing models with different spin in
Figure 4.  The X-ray flux therefore increases with $a_*$ because
$\Theta_e$ at the ISCO increases.

(5) The dependence on $T_i/T_e$ is mainly due to synchrotron
self-absorption, which is strongest at high inclination.  For example,
because the $i = 85^{\deg}$, $T_i/T_e = 10$ model is optically thick at
$230 GHz$ the emission is produced in a synchrotron photosphere well
outside $r_{ISCO}$.  The typical radius of the synchrotron photosphere
ranges between 15 $G M/c^2$ for low spin models ($a_*=0.5,0.75$) and 8
$G M/c^2$ for high spin models ($a_*> 0.75$). The $230 GHz$ flux can
then be produced only with large ${\mathcal M}$; as ${\mathcal M}$
increases the optically thin flux in the NIR increases due to increasing
density and field strength. The scattered spectrum also depends on
$T_i/T_e$ since the energy boost per scattering is $\sim 16\Theta_e^2
\propto 1/(T_i/T_e)^2$.

(6) The inclination dependence is a relativistic effect.  ${\mathcal M}$
is nearly independent of $i$ (it varies by $\sim 10\%$, except for
$T_i/T_e = 10$, which due to optical depth effects has much larger
variation), so models with different inclination are nearly identical.
Nevertheless the X-ray flux varies dramatically with $i$, increasing by
almost 2 orders of magnitude from $i = 5\deg$ to $i = 85^{\deg}$.  This
occurs because Compton scattered photons are beamed forward parallel to
the orbital motion of the disk gas.  The variation of mm flux with $i$
is due to self-absorption.  The mm flux reflects the temperature and
size of the synchrotron photosphere.  At lower $i$ the visible
synchrotron photosphere is hotter than at high $i$.

\section{Future prospects}

We have made a number of approximations that will be removed in future
models of Sgr A*.  We plan to (1) add cooling, (2) run 3D (rather than
axisymmetric) models, (3) model a larger range of radii, (4) include a
population of nonthermal electrons, (5) eliminate the fast-light
approximation by doing fully time-dependent radiative transfer.

\acknowledgements 

This work was supported by the National Science Foundation under grants
AST 00-93091, PHY 02-05155, and AST 07-09246, by NASA grant NNX10AD03G,
through TeraGrid resources provided by NCSA and TACC, and by a Richard
and Margaret Romano Professorial scholarship, a Sony faculty fellowship,
and a University Scholar appointment to CFG.  


\end{document}